\documentclass[prl]{revtex4}
\usepackage{bm}
\usepackage{epsfig}
\usepackage{amsmath}
\begin{document}
\title{Buckled nano rod - a two state system and its dynamics using system plus reservoir model}
\author{Aniruddha Chakraborty,$^{1,2}$}
\affiliation{\emph{$^{1}$Department of Inorganic and Physical
Chemistry, Indian Institute of Science, Bangalore, 560012, India,}\\
\emph{$^{2}$ Jawaharlal Nehru Center for Advanced Scientific
Research, Bangalore, 560064, India.}}
\date{\today}

\begin{abstract}
\noindent We consider a suspended elastic rod under longitudinal
compression. The compression can be used to adjust potential
energy for transverse displacements from harmonic to double well
regime. As compressional strain is increased to the buckling
instability, the frequency of fundamental vibrational mode drops
continuously to zero (first buckling instability). As one tunes
the separation between ends of a rod, the system remains stable
beyond the instability and develops a double well potential for
transverse motion. The two minima in potential energy curve
describe two possible buckled states at a particular strain. From
one buckled state it can go over to the other by thermal
fluctuations or quantum tunnelling. Using a continuum approach and
transition state theory (TST) one can calculate the rate of
conversion from one state to other. Saddle point for the change
from one state to other is the straight rod configuration. 
The rate, however, diverges at the
second buckling instability. At this point, the straight rod
configuration, which was a saddle till then, becomes hill top and
two new saddles are generated. The new saddles have bent
configurations and as rod goes through further instabilities, they
remain stable and the rate calculated according to harmonic
approximation around saddle point remains finite.
In our earlier paper classical rate calculation
including friction has been carried out [J. Comput. Theor. Nanosci. {\bf 4} (2007) {\it 1}], by assuming that each
segment of the rod is coupled to its own collection of harmonic
oscillators - our rate expression is well behaved through the second
buckling instability. In this paper we have extended our method to calculate quantum rate using the same system plus reservoir model. We find that friction lowers the rate of conversion.

\end{abstract}

\maketitle

\section{Introduction}
\begin{figure}[h]
\centering
\epsfig{file=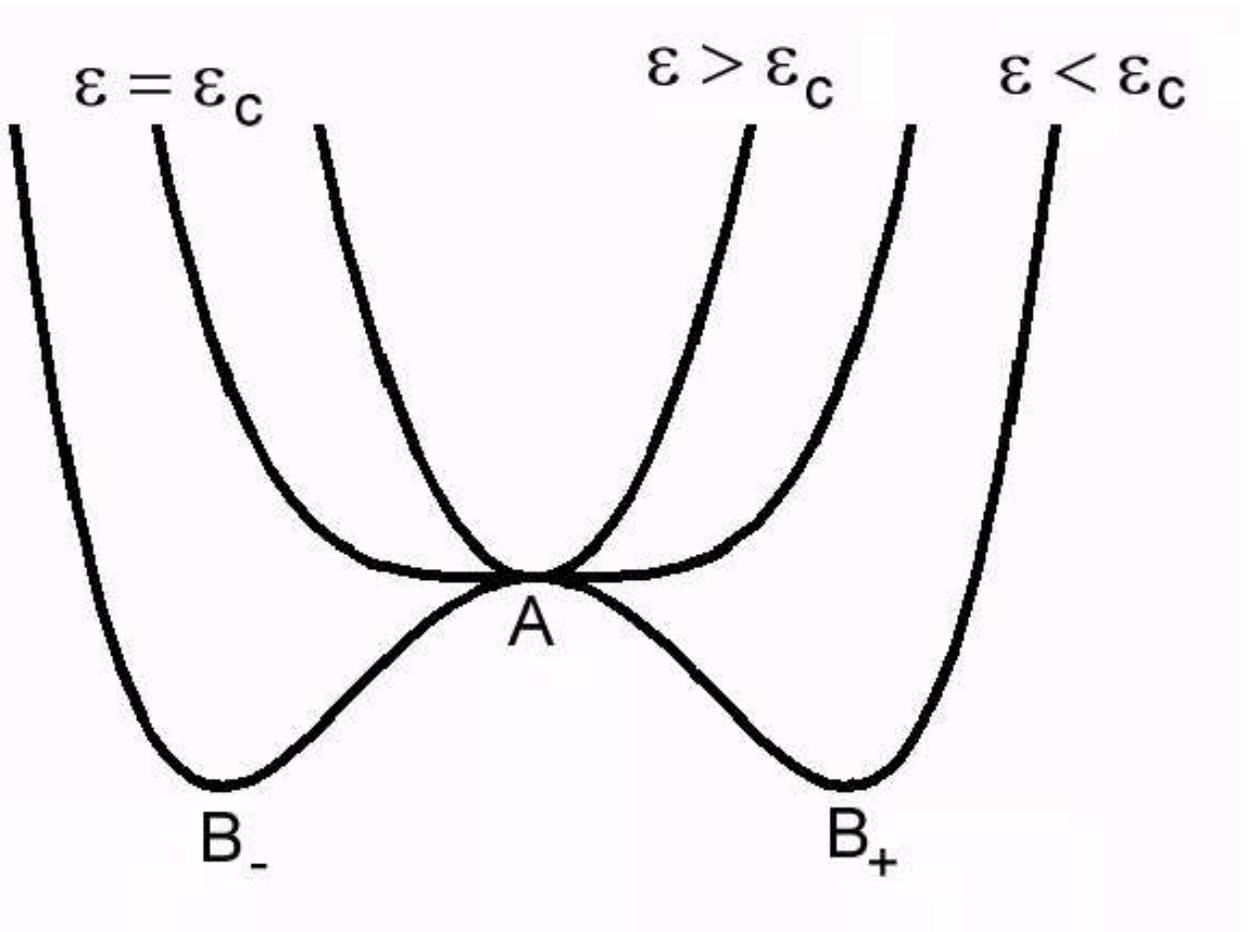,width=1\linewidth} \\
\caption{Potential energy $V$ as a function of the fundamental
mode displacement $Y$. The shape of the potential energy is
harmonic for $ \varepsilon >\varepsilon _c$, quartic for
$\varepsilon =\varepsilon _c\equiv $ critical strain ($\varepsilon
_c<0$) and a double well for $\varepsilon <\varepsilon _c$.}
\label{figpotential}
\end{figure}

\noindent Considerable attention has recently been paid to
two-state nano-mechanical systems  \cite{Roukes, Craighead, Park,
Erbe, ClelandAPL, Rueckes} and the possibility of observing
quantum effects in them. In the experiments of Rueckes {\it et
al.} \cite{Rueckes} crossed carbon nano-tubes were suspended
between supports and the suspended element was electrostatically
flexed between two states. Roukes {\it et al.} \cite{Roukes}
propose to use an electrostatically flexed cantilever to explore
the possibility of macroscopic quantum tunnelling in a
nano-mechanical system. Carr {\it et al.} \cite{Carr,ClelandAPL}
suggest using the two buckled states of a nanorod as the two
states and investigate the possibility of observing quantum
effects. Here we consider a suspended elastic rod of rectangular
cross section under longitudinal compression. The compression is
used to adjust the potential energy for transverse displacements
from the harmonic to the double well regime as shown in the
Fig.\ref{figpotential} As the compressional strain is increased to
the buckling instability \cite{Euler}, the frequency of the
fundamental vibrational mode drops continuously to zero. Beyond
the instability, the system has a double well potential for the
transverse motion.
\noindent The two minima in the potential energy curve describe
the two possible buckled states at a particular strain \cite{Carr}
and the system can change from one state to the other, under
thermal fluctuations or quantum tunneling. In our earlier publications \cite{Ani,Sayan} we have used
multidimensional transition state theory (TST) to derive
expressions for the transition rate from one potential well to the
other. We now include the effect of friction on the reaction
rate. For this, we follow the procedure \cite{Weiss} used for the
study of barrier crossing and other dynamical problems in the
presence of friction.

\begin{figure}[h]
\centering
\epsfig{file=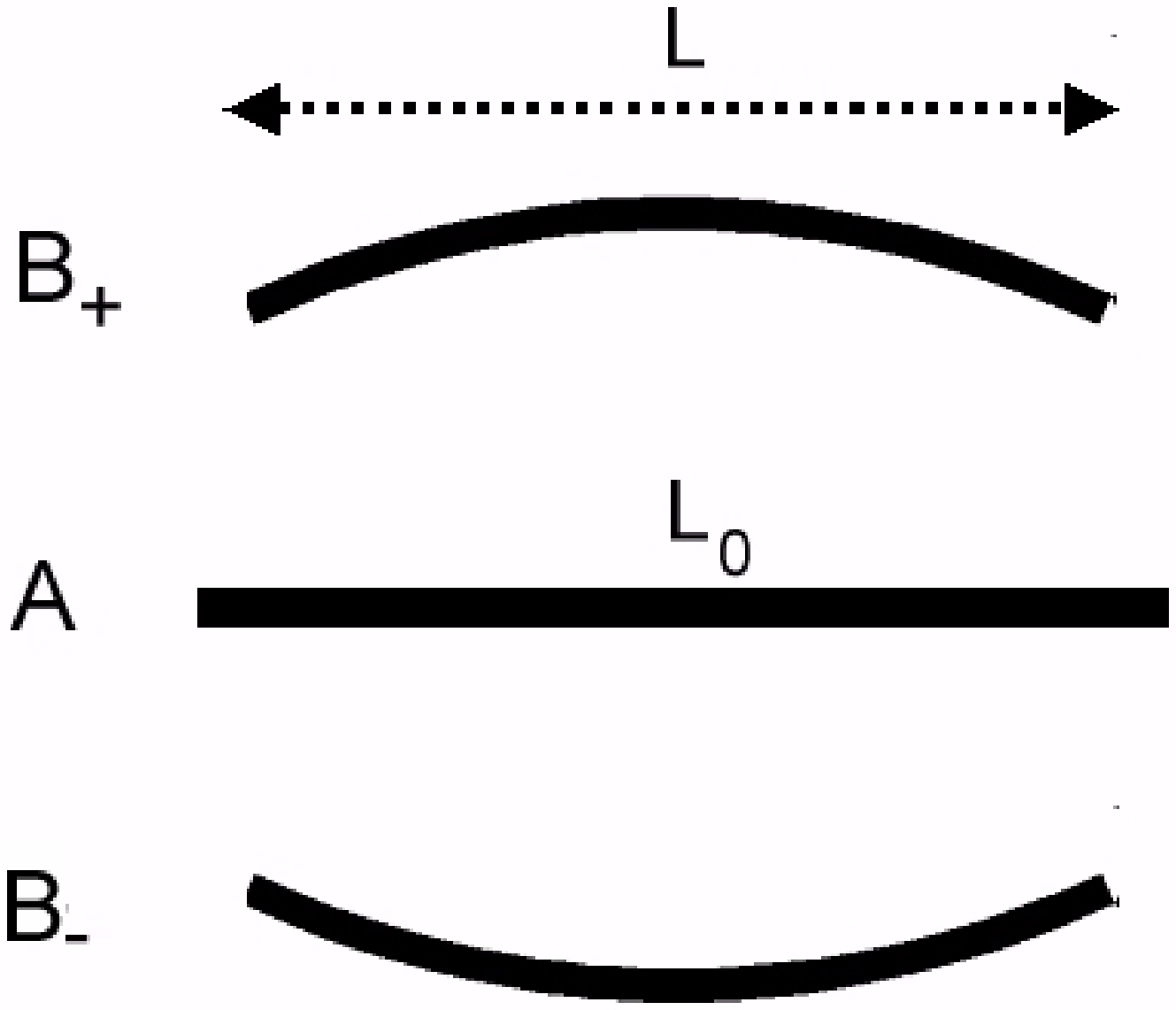,width=1\linewidth} \\
\caption{The rod under compression: The central figure (A) shows
the uncompressed rod of length $L_{0}$. On compressing to length
$L$, the rod buckle, either to $B_{-}$ or to $B_{+}$.}
\label{figB2}
\end{figure}

\section{The Model}

\noindent We consider the normal modes and associated quantum
properties of an elastic rectangular rod of length $L$, width $w$
and thickness $d$ (satisfying $L>>w>>d$) \cite{Carr, Wybourne,
CarrAPL, Lawrence}. We assume that $d$ is smaller than $w$ so that
transverse displacements $y(x)$ only occur in the ``$d$ ''
direction. $F$ is the linear modulus (energy per unit length) of
the rod and is related to the
elastic modulus $Q$ of the material by $F=Q$ $w$ $d$. The bending moment $%
\kappa $ is given by $\kappa ^2=\frac{d^2}{12}$ for a rod of
rectangular cross section and $\mu =\frac mL$ is the mass per unit
length. We take the length of the rod (uncompressed) to be $L_0$.
As in the Fig. \ref {figB2}, we apply compression on the two ends,
reducing the separation between the two to $L$. Then if $y(x)$ denotes the displacement of the rod in
the `$d$' direction, the total potential energy is given by \cite{Sayan,Ani}
\begin{equation}
\label{B125}V[y(x)]=\frac 12\int\limits_0^Ldx(F\kappa ^2(y^{\prime
\prime })^2+F\varepsilon (y^{\prime })^2)+\frac F{8L_0}(\int\limits_0^Ldx(y^{\prime })^2)^2+\frac
{FL_0}{2}\varepsilon ^2.
\end{equation}
In the above $\varepsilon =\frac{(L-L_0)}{L_0}$ is the strain,
negative if compressive. We imagine that each segment of the rod
(labelled by $x$) is coupled to a collection  of harmonic
oscillators which form the reservoir. Each section has its own
independent collection of harmonic oscillators. So the total
system, rod plus the reservoir has the energy \cite{Sayan}
\begin{align}
\label{B87}&H=\frac 12\int\limits_0^Ldx\ \mu
\stackrel{.}{y}^2+V[y]+\frac 12\sum_{\alpha=1}^N \int\limits_0^Ldx\nonumber\\
&\times \left\{ m_\alpha (x)\stackrel{.}{\xi }_\alpha ^2
          +m_\alpha (x)\omega _\alpha ^2(x)\left( \xi _\alpha
(x)-\frac{c_\alpha (x)y(x)}{m_\alpha (x)\omega _\alpha
^2(x)}\right) ^2\right\}.
\end{align}
\noindent In the above, $\xi _\alpha (x)$ denotes the position of $(\alpha ,x)$-th harmonic oscillator of the reservoir, coupled to $y(x)$, the displacement of the rod at location $x$. It has a frequency $\omega _\alpha(x)$ and mass $m_\alpha (x)$, $c_\alpha (x)$ determines the coupling of the $\xi _\alpha (x)$ to the $y(x)$. The way, the coupling has been chosen, the barrier height remains unchanged. Here $\alpha $ varies over the collection of harmonic oscillators, and we let $\alpha $ vary from $1$ to $N$. 

\section{Quantum transition state theory in presence of friction}
\noindent To derive the quantum rate we use the following methodology. The Hamiltonian given in Eq. (\ref{B87}) may be
treated as a quantum Hamiltonian. For a finite discrete set of oscillators one may evaluate the quantum rate using quantum
transition state theory under harmonic approximations. The quantum transition state theory rate expression is given by
\cite{PollakCPL,PollakPRA}
\begin{equation}
\label{B108}R_{quantum}^{fric}= \frac{kT}h \left(\frac{\hbar
\beta\Omega^{fric}}{2\sin (\frac 12\hbar \beta
\Omega^{fric})}\right)\frac{Q^{\ddagger }}Qe^{-\beta E_{act}}.
\end{equation}
\noindent In the above $Q^{\ddagger}$ and $Q$ are the partition
function at the saddle point and at the reactant, and
$\Omega^{fric}$ is the frequency of the `unstable mode'.
Following Pollak \cite{PollakCPL}, the quantum partition function
at the saddle point
\begin{align}
\label{B109} Q^{\ddagger }= \prod\limits_{n=2}^\infty
                            &\left(\frac 1{2\sinh (\frac 12\hbar \beta \chi_n^{\ddagger
})}\prod\limits_{j=1}^N\frac 1{2\sinh (\frac 12\hbar \beta \lambda
_{n,j}^{\ddagger })}\right).
\end{align}
\noindent At the saddle point we have $N \times N-1$ real
oscillators and one imaginary oscillator. $\lambda_{nj}^{\ddagger }$ is the `bath' frequency and
$\chi_n^{\ddagger }$ is the `system' frequency at the saddle point. Note the well
known divergence of $Q^{\ddagger }$ at low temperatures. The
quantum partition function at the reactant geometry
\begin{equation}
\label{B110} Q=\prod\limits_{n=1}^\infty \left(\frac 1{2\sinh
(\frac 12\hbar \beta \chi _n)}\prod\limits_{j=1}^N\frac 1{2\sinh
(\frac 12\hbar \beta \lambda _{n,j})}\right).
\end{equation}
\noindent In the above $\lambda _{n,j}$ is the `bath' frequency
and $\chi_n$ is the `system' frequency at the reactant geometry.

\section{The Extrema of the functional $V[y(x)]$}

\noindent To find the equilibrium state, we extremise potential
energy functional with respect to $y(x)$. For this we put
$\frac{\delta V[y(x)]}{\delta y(x)}=0$. This leads to the
differential equation
\begin{equation}
\label{C126}F\kappa ^2\frac{\partial ^4y}{\partial x^4}-[F\varepsilon \frac{%
\partial ^2y}{\partial x^2}+\frac F{2L_0}(\int\limits_0^Ldx(y^{\prime
}(x))^2)\frac{\partial ^2y}{\partial x^2}]=0
\end{equation}
\noindent and the hinged end points have boundary conditions
$y(0)=y(L)=0$ and $y^{\prime \prime }(0)=y^{\prime \prime}(L)=0.$
The only solution to Eq. (\ref{C126}) is $y(x)=0$ if $\varepsilon
>\varepsilon _c,$ where $\varepsilon _c=-\frac{\kappa ^2\pi
^2}{L^2}.$ But if $\varepsilon <\varepsilon _c$ two more solutions
are possible. They are given by
\begin{equation}
\label{C127}y(x)=\pm A\sqrt{\frac 2L}\sin (\frac \pi Lx).
\end{equation}
\noindent On substituting Eq. (\ref{C127}) into Eq. (\ref{C126}), we
find a nonlinear equation for $A$, which on solution gives
\begin{equation}
\label{B28}A=\sqrt{\frac{2L_0L^2}{
\pi ^2}(\varepsilon
_c-\varepsilon )}.
\end{equation}
\noindent These two are the buckled states which are minima of the
potential energy surface for $\varepsilon <\varepsilon _c$, as all
the normal modes around this are stable. The solution $y(x)=0$ is
now a saddle point (see next section for details).

\section{Normal modes of the system Hamiltonian}

\noindent We find the potential energy for the buckled states
$E_{b}=-\frac{FL_0}2(\varepsilon -\varepsilon _c)^2$. $y(x)=0$ is
the straight rod configuration and is the saddle point for
transition from one buckled state to the other. Its potential
energy $E_{Saddle}^{Linear}$ is zero. One can calculate the
barrier height for the process of going from one buckled state to
the other as
\begin{equation}
\label{C29}\Delta E_{Barrier}^{Linear}=\frac{FL_0} 2(\varepsilon
-\varepsilon _c)^2.
\end{equation}
\noindent The kinetic energy of the rod is $\frac\mu2
\int_{0}^{L} y_{t}^{2}dx$, where $\mu =\frac{m}{L_{0}}$ is the
mass per unit length. We now look at small amplitude vibrations
around each extremum $y_{ext}(x)$, which could be the buckled or
the saddle point. One gets 
\begin{equation}
\label{C30}\mu \frac{\partial ^2\delta y(x,t)}{\partial t^2}%
=-\int\limits_0^Ldx_1\left( \frac{\delta^2 V[y(x)]}{\delta y(x_1)\delta y(x)}%
\right) _{y_{ext}(x_1)}\delta y(x_1,t).
\end{equation}
\noindent On evaluation we get
\begin{equation}
\label{C31}\mu \frac{\partial ^2\delta y(x,t)}{\partial t^2}=-{\cal L}%
_{ext}\delta y(x,t),
\end{equation}
\noindent where the operator ${\cal L}_{ext}$ is defined by
\begin{align}
\label{C32} {\cal L}_{ext}\delta y(x,t)=&F\kappa ^2\delta
y^{\prime \prime \prime \prime }(x,t)-F\varepsilon \delta
y^{\prime \prime }(x,t)-\frac F{2L_0}(\int\limits_0^Ldx(y_{ext}^{\prime }(x))^2)\delta
y^{\prime
\prime }(x,t) \nonumber\\
                                        &+\frac F{L_0}y_{ext}^{\prime \prime
}(x)(\int\limits_0^Ldx_1\;y_{ext}^{\prime \prime }(x_1)\delta
y(x_1,t)).
\end{align}
\noindent Using the boundary conditions for the hinged end points,
we find the normal modes of the rod, $\delta y(x,t)=\delta
y_n(x)e^{i\omega _nt}$. At the saddle point, $y_{ext}(x)=$
$y_0(x)=0$. Using this in Eq. (\ref {C32}), we obtain
\begin{equation}
\label{C33}\delta y_n(x)=A_n \sqrt{\frac2L}\sin (\frac{n\pi }Lx),
\end{equation}
\noindent with n=1,2,3.... The normal mode frequencies at the
saddle point are given by ($\ddagger$ is used to denote the saddle
point)
\begin{equation}
\label{C34}\,\omega _{Linear,n}^{\ddagger }=\omega
_0\;n\,\sqrt{n^2-\frac \varepsilon {\varepsilon _c}}.
\end{equation}
\noindent $\omega _0=\frac{\pi ^2\kappa }{L^2}\sqrt{\frac F\mu }$.
$n=1$ is the unstable mode and it has the imaginary frequency
$\omega _1=i\Omega _{Linear}$, where
\begin{equation}
\label{C35}\Omega _{Linear}=\omega _0\;\,\sqrt{\frac \varepsilon
{\varepsilon _c}-1}.
\end{equation}
\noindent
For small amplitude vibrations around the buckled state, one has to put $%
y_{ext}(x)=A \sqrt{\frac 2L}\sin (\frac \pi Lx)$. The normal modes
are the same as in the Eq. (\ref{B33}), but the normal mode
frequencies are different. They are
\begin{equation}
\label{C36}\omega _n=\omega _0n\sqrt{n^2-1}\qquad for\quad n>1,
\end{equation}
\noindent while
\begin{equation}
\label{C36a}\omega _1=\omega _0\sqrt{2\left( \frac \varepsilon
{\varepsilon _c}-1\right) }.
\end{equation}

\subsection{The Rate near first buckling instability}
\noindent The reaction rate using quantum transition state theory may be written as \cite{PollakCPL}
\begin{equation}
\label{B111} R_{quantum,f}^{fric}=\rho_f(\frac{\omega _1}{2\pi
})(\frac{\Omega^{fric}_{Linear}}{\Omega_{Linear}})\prod\limits_{n=2}^\infty
\frac{\omega _n}{\omega _{Linear,n}^{\ddagger }}e^{-\beta \Delta E_{Barrier}^{Linear}},
\end{equation}
\noindent where, we have used the following expression for $\Omega_{Linear}^{fric} $ \cite{Ani}
\begin{equation}
\label{C98}\Omega_{Linear}^{fric}=-\frac \gamma 2+\sqrt{\left( \frac
\gamma 2\right) ^2+\omega_0^2\left( \frac{ \varepsilon
}{\varepsilon _c} -1\right) }
\end{equation}
and the expression for $\rho$ is given by
\begin{align}
\label{B111a}  \rho_f =\frac{\Omega_{Linear}}{\omega_1}
\prod\limits_{n=2}^\infty \left( \frac{\omega_{Linear,n}^{\ddagger}}{\omega _n}\right) \frac{\sinh (\frac 12\hbar \beta \chi_1)\prod\limits_{j=1}^N\frac 1{2\sinh (\frac 12\hbar \beta \lambda
_{n,j}^{\ddagger })}}{\sin (\frac 12\hbar \beta \Omega_{Linear}^{fric
})\prod\limits_{j=1}^N\frac 1{2\sinh (\frac 12\hbar \beta \lambda
_{n,j})}}\nonumber \\ \frac{\prod\limits_{n=2}^\infty \left( \frac
1{2\sinh (\frac 12\hbar \beta \chi _n^{\ddagger
})}\prod\limits_{j=1}^N\frac 1{2\sinh (\frac 12\hbar \beta \lambda
_{n,j}^{\ddagger })}\right) }{\prod\limits_{n=2}^\infty \left( \frac
1{2\sinh (\frac 12\hbar \beta \chi_n)}\prod\limits_{j=1}^N\frac
1{2\sinh (\frac 12\hbar \beta \lambda _{n,j})}\right) }.
\end{align}
\noindent In the above $\omega_n$ and $\omega_{Linear,n}^{\ddagger}$
represents frequencies of normal modes of the system at reactant
geometry and at the saddle point respectively in absence of bath. In case when the system is coupled to the bath $\lambda_{n,j}^{\ddagger }$ is the new `bath' frequency and $\chi_n^{\ddagger }$ is the new `system' frequency at the saddle point, also $\lambda _{n,j}$ is the new `bath' frequency and $\chi_n$ is the new `system' frequency at the reactant geometry. To evaluate $\rho$, we use the following two identities \cite{Ryzhik}
\begin{equation}
\label{B112} \sinh (x)=x\prod\limits_{k=1}^\infty
(1+\frac{x^2}{k^2\pi ^2})
\end{equation}
\noindent and
\begin{equation}
\label{B113} \sin(x)=x\prod\limits_{k=1}^\infty
(1-\frac{x^2}{k^2\pi ^2}).
\end{equation}
\noindent Following Pallak \cite{PollakCPL} one can prove the following identity
\begin{align}
\label{B104a}&({-{\Omega^{fric}_{Linear}}^2+\epsilon^2})\prod\limits_{i=1}^N
({\lambda_{1,i}^{\ddagger 2}+\epsilon^2})\times\prod\limits_{n=2}^\infty \left( ({\chi _n^{\ddagger
2}+\epsilon^2})\prod\limits_{j=1}^N ({\lambda _{n,j}^{\ddagger
2}+\epsilon^2})\right) \nonumber \\&  = \left( {-\Omega_{Linear}
^2+\epsilon^2+\epsilon \gamma}\right)\prod\limits_{\alpha=1}^N({\omega
_\alpha^{2}+\epsilon^2}) \times
\prod\limits_{n=2}^\infty \left( ({\omega_{Linear,n}^{\ddagger
2}+\epsilon^2})\prod\limits_{\alpha=1}^N({\omega
_\alpha^{2}+\epsilon^2})\right),
\end{align}
\noindent where $\epsilon$ is an arbitrary number. Also
\begin{align}
\label{B104b}&({{\chi
_1}^2+\epsilon^2})\prod\limits_{i=1}^N ({\lambda _{1,i}^{
2}+\epsilon^2})\times\prod\limits_{n=2}^\infty \left( ({\chi _n^{
2}+\epsilon^2})\prod\limits_{j=1}^N ({\lambda _{n,j}^{
2}+\epsilon^2})\right) \nonumber \\&  = \left( {\omega_1
^2+\epsilon^2+\epsilon \gamma}\right)\prod\limits_{\alpha=1}^N({\omega
_\alpha^{2}+\epsilon^2}) \times
\prod\limits_{n=2}^\infty \left( ({\omega _n^{
2}+\epsilon^2})\prod\limits_{\alpha=1}^N({\omega
_\alpha^{2}+\epsilon^2})\right).
\end{align}
\noindent We use the notation $\nu=\frac{2 \pi}{\hbar \beta}$.
Insertion of Eq. (\ref{B113}) into Eq. (\ref{B112})
and use of Eq. (\ref{B104a}) and  Eq. (\ref{B104b}) with
$\epsilon=0$ gives
\begin{align}
\label{B114} \rho_f = & \left( \prod\limits_{k=1}^\infty
\frac{(k^2\nu ^2+\chi _1^2)}{(k^2\nu ^2-\Omega^{2}_{fric})}\right)
\prod\limits_{i=1}^N \prod\limits_{k=1}^\infty \left(
\frac{(k^2\nu ^2+\lambda_{1,i}^2)}{(k^2\nu ^2+\lambda_{1,i}^{\ddagger
2})}\right) \nonumber \\& \prod\limits_{n=2}^\infty \left(
\left(\prod\limits_{k=1}^\infty \frac{(k^2\nu ^2+\chi
_n^2)}{(k^2\nu ^2+\chi _n^{\ddagger 2})}\right)
\prod\limits_{i=1}^N \prod\limits_{k=1}^\infty \left(
\frac{(k^2\nu ^2+\lambda _{1,i}^2)}{(k^2\nu ^2+\lambda
_{1,i}^{\ddagger 2})}\right)\right).
\end{align}
\noindent Interchanging the order of the products and using ratio
of Eq. (\ref{B104a}) and Eq. (\ref{B104b})with the identification
$\epsilon=k \nu$ gives the desired result.
\begin{align}
\label{B115}\rho_f =\left( \prod\limits_{k=1}^\infty \frac{(k^2\nu
^2+\omega _1^2+\nu k\gamma )}{(k^2\nu ^2-\Omega^2_{fric}+\nu k\gamma
)}\right)  \nonumber\\ \prod\limits_{n=2}^\infty \left(
\prod\limits_{k=1}^\infty \frac{(k^2\nu ^2+\omega _n^2+\nu k\gamma
)}{(k^2\nu ^2+\omega_{Linear,n}^{\ddagger 2}+\nu k\gamma )}\right).
\end{align}
\noindent On simplification we get
\begin{align}
\label{B115ab} \rho_f= \frac{\Gamma \left( \frac{\gamma +2\nu
-\sqrt{\gamma ^2+4\Omega^2_{Linear}}}{2\nu }\right) \Gamma \left(
\frac{\gamma +2\nu +\sqrt{\gamma ^2+4\Omega^2_{Linear}}}{2\nu }\right)
}{\Gamma \left( \frac{\gamma +2\nu -\sqrt{\gamma ^2-4(\omega
_1)^2}}{2\nu }\right) \Gamma \left( \frac{\gamma +2\nu
+\sqrt{\gamma ^2-4(\omega _1)^2}}{2\nu }\right) }\nonumber\\\times
\prod\limits_{n=2}^\infty \frac{\Gamma \left( \frac{\gamma +2\nu
-\sqrt{\gamma ^2-4(\omega_{Linear,n}^{\ddagger })^2}}{2\nu }\right) \Gamma
\left( \frac{\gamma +2\nu +\sqrt{\gamma ^2-4(\omega_{Linear,n}^{\ddagger
})^2}}{2\nu }\right) }{\Gamma \left( \frac{\gamma +2\nu
-\sqrt{\gamma ^2-4(\omega _n)^2}}{2\nu }\right) \Gamma \left(
\frac{\gamma +2\nu +\sqrt{\gamma ^2-4(\omega _n)^2}}{2\nu }\right)
}
\end{align}

\section{Beyond the Second Buckling Instability}

\noindent As $\sqrt{\frac \varepsilon {\varepsilon _c}}\rightarrow
2$, $\omega _{Linear,2}^{\ddagger }\rightarrow 0$ and the reaction
rate diverges [see Eq. (18)]. So the rate expression in Eq. (18) is valid only if one is not too
near $\sqrt{\frac \varepsilon {\varepsilon _c}}=2$. This is due to
the setting in of the second buckling instability. As the rod is
compressed, first the mode $A_1\sqrt{\frac 2L} \sin(\frac \pi Lx)$
becomes unstable and this is the first buckling instability and
the rod buckles as a result of this. The length at which this
occurs shall be denoted by $L_f$. If one supposes that the rod is
compressed further keeping the straight rod configuration, then at
a length $L_s$, the mode $A_2\sqrt{\frac 2L}\sin(\frac{2\pi }Lx)$
too would become unstable and this is the second buckling
instability. What happens here is a reaction path bifurcation for
the crossing from one buckled state to the other and is very
interesting. For $\varepsilon > 4\varepsilon _c$, there is only
one saddle point but for $\varepsilon < 4\varepsilon _c$, this
saddle point bifurcates into two and consequently the calculation
of rate near the bifurcation is a challenging problem. In a similar fashion one can have the third
instability at a length $L_t$ etc. but these present no problem as
far as rate calculation is concerned (see below). In order to
analyze the rate near and beyond the second buckling instability,
we proceed as follows. We assume that the Eq. (12) has
solutions of the form
\begin{equation}
\label{C40}y_0(x)=A_1\sqrt{\frac 2L}\sin(\frac \pi Lx)+A_2\sqrt{\frac 2L}\sin(%
\frac{2\pi }Lx).
\end{equation}
\noindent Using this, the elastic potential energy is given by
\begin{align}
\label{C41}V[A_1,A_2]=& \frac{F\pi
^4(A_1^2+4A_2^2)^2}{8L^4L_0}+\frac{F\pi ^2A_1^2(\varepsilon
-\varepsilon _c)}{2L^2}\nonumber\\&+\frac{2F\pi
^2A_2^2(\varepsilon -4\varepsilon _c)}{L^2}.
\end{align}
\noindent The two buckling instabilities are clearly evident from
this expression - as each the coefficient of $A_1^2$ or $A_2^2$
changes sign from positive to negative. Finding the extrema of
this potential leads to the following solutions for ($A_1$, $A_2$)

1. ($0$, $0$): this is the straight rod configuration. Between
first and second buckling (i.e. $L_s<L<L_f$ ), this is the saddle point. But
after the second buckling, it is no longer a saddle, but it
becomes a hill top. It has an energy $E_{hilltop}=0$.

2. ($\pm \frac 2\pi \sqrt{LL_0(\varepsilon _c-\varepsilon )}$,
$0$): These
are the buckled states and both of them have the same energy $E_b=-\frac{FL_0%
}2(\varepsilon -\varepsilon _c)^2$.

3. ($0$, $\pm \frac 1\pi \sqrt{LL_0(4\varepsilon _c-\varepsilon
)}$): These are the two new saddle points that arise from the
bifurcation of the one that existed for $4\varepsilon _c>
\varepsilon$. These two have the same energy
$E_{saddle}^{Bent}=-\frac{FL_0}2(\varepsilon -4\varepsilon _c)^2$.
At these saddle points, the rod has a bent (S-shaped) geometry. Beyond the second buckling instability, the barrier height is given by
\begin{equation}
\label{C42} \Delta E_{Barrier}^{Bent}=-\frac{3FL_0\varepsilon
_c}2(-2\varepsilon +5\varepsilon _c).
\end{equation}
\noindent Beyond the second buckling instability, away from the
instability, one can do a normal mode analysis near the vicinity
of the new saddles - there are two of them, both making identical
contributions to the reaction rate. Near the buckled state, the
normal modes have the frequencies given in the Eq. (\ref{C36}) and
Eq. (\ref{C36a}), while near the saddle, the frequencies are given
by
\begin{equation}
\label{C69}\omega _{Bent,1}^{\ddagger }=\Omega _{Bent},
\end{equation}
\noindent
\begin{equation}
\label{C70}\omega _{Bent,2}^{\ddagger }=\omega _0\sqrt{8(\frac
\varepsilon {\varepsilon _c}-4)},
\end{equation}
\noindent
\begin{equation}
\label{C71}\omega _{Bent,n}^{\ddagger }=\omega _0n\sqrt{n^2-4},
\end{equation}
\noindent for $n>2$. In the above $\omega _{Bent,1}^{\ddagger }$
and has an imaginary frequency with
\begin{equation}
\label{C69a}\Omega _{Bent}=\sqrt{3}\omega _0.
\end{equation}
Now the quantum rate beyond the second buckling instability can be calculated taking the saddle to be the bent configuration. 
\begin{equation}
\label{C111} R_{quantum,s}^{fric}=2 \times \rho_s(\frac{\omega _1}{2\pi
})(\frac{\Omega^{fric}_{Bent}}{\Omega_{Bent}})\prod\limits_{n=2}^\infty
\frac{\omega _n}{\omega _{Bent,n}^{\ddagger }}e^{-\beta \Delta E_{Barrier}^{Bent}},
\end{equation}
\noindent where, we have used the following expression for $\Omega_{Bent}^{fric} $ \cite{Ani}
\begin{equation}
\label{C198}\Omega_{Bent}^{fric}=-\frac \gamma 2+\sqrt{\left( \frac
\gamma 2\right) ^2+ 3 \omega_0^2}
\end{equation}
and the expression for $\rho_s$ is given by
\begin{align}
\label{B215ab} \rho_s= \frac{\Gamma \left( \frac{\gamma +2\nu
-\sqrt{\gamma ^2+4\Omega^2_{Bent}}}{2\nu }\right) \Gamma \left(
\frac{\gamma +2\nu +\sqrt{\gamma ^2+4\Omega^2_{Bent}}}{2\nu }\right)
}{\Gamma \left( \frac{\gamma +2\nu -\sqrt{\gamma ^2-4(\omega
_1)^2}}{2\nu }\right) \Gamma \left( \frac{\gamma +2\nu
+\sqrt{\gamma ^2-4(\omega _1)^2}}{2\nu }\right) }\nonumber\\\times
\prod\limits_{n=2}^\infty \frac{\Gamma \left( \frac{\gamma +2\nu
-\sqrt{\gamma ^2-4(\omega_{Bent,n}^{\ddagger })^2}}{2\nu }\right) \Gamma
\left( \frac{\gamma +2\nu +\sqrt{\gamma ^2-4(\omega_{Bent,n}^{\ddagger
})^2}}{2\nu }\right) }{\Gamma \left( \frac{\gamma +2\nu
-\sqrt{\gamma ^2-4(\omega _n)^2}}{2\nu }\right) \Gamma \left(
\frac{\gamma +2\nu +\sqrt{\gamma ^2-4(\omega _n)^2}}{2\nu }\right)
}
\end{align}
We have multiplied the quantum rate by a factor of 2 to account for the fact that there are two saddles of equal energy. It is interesting that the normal modes for this saddle retain their stability, irrespective of what the compression is. The first mode is always unstable and other modes always stable for all values of $\epsilon$. Therefore, this rate expression is valid for all values of $\epsilon < 4 \epsilon_c$ - that is even through the third buckling instability.

\subsection{The rate near second buckling instability}

\noindent Near the second buckling instability ($\sqrt{\frac
\varepsilon {\varepsilon _c}}\rightarrow 2$), both $\omega
_{Linear,2}^{\ddagger }$  and $\omega_{Bent,2}^{\ddagger }$ vanishes, causing the rate to diverge
[see Eq. (18) and Eq. (35)]. The cure for this divergence is to go beyond the harmonic approximations for the first two modes. Our discussion here follows that of Weiss
\cite{Weiss}. For the model described by the Hamiltonian
\begin{align}
\label{B115aa}&H=\frac 12\int\limits_0^Ldx\ \mu
\stackrel{.}{y}^2+V[y]+\frac 12\sum_\alpha \int\limits_0^Ldx\nonumber\\
&\times \left\{ m_\alpha (x)\stackrel{.}{\xi }_\alpha ^2
          +m_\alpha (x)\omega _\alpha ^2(x)\left( \xi _\alpha
(x)-\frac{c_\alpha (x)y(x)}{m_\alpha (x)\omega _\alpha
^2(x)}\right) ^2\right\}.
\end{align}
\noindent We note that $m_\alpha (x)$ is independent of $x$ and
may be written as $m_\alpha $, $\omega _\alpha ^2(x)$ too, as well
as $c_\alpha (x)$. So we can write the above expression as
\begin{align}
\label{B115bb} & H =\frac 12\int\limits_0^Ldx \mu
\stackrel{.}{y}^2 +V[y]+ \nonumber\\ &\frac 12\sum\limits_\alpha
\int dx\left\{ m_\alpha \stackrel{.}{\xi }_\alpha ^2(x)+m_\alpha
\omega _\alpha ^2\left( \xi _\alpha (x)-\frac{c_\alpha
y(x)}{m_\alpha \omega _\alpha ^2}\right) ^2\right\}.
\end{align}
Now let
\begin{equation}
\label{C1} \xi _\alpha (x)=\sum\limits_n \sqrt{\frac 2L}\xi
_{\alpha ,n}\sin (\frac{n\pi x}L)
\end{equation}
\noindent and
\begin{equation}
\label{C2} y(x)=\sum\limits_n \sqrt{\frac 2L}A_n\sin (\frac{n\pi
x}L).
\end{equation}
\noindent Then,
\begin{align}
\label{C3} & H=\frac 12\mu \sum\limits_n
\stackrel{.}{y_n}^2+V[A_1,A_2,....A_n]\nonumber\\ &+\frac
12\sum\limits_{\alpha ,n}\left\{ m_\alpha \stackrel{.}{\xi
}_{\alpha ,n}^2+m_\alpha \omega _\alpha ^2\left( \xi _{\alpha
,n}-\frac{c_\alpha A_n}{m_\alpha \omega _\alpha ^2}\right)
^2\right\},
\end{align}
\noindent which decouples all the modes. The Hamiltonian for the
first two modes coupled with bath is given by
\begin{align}
\label{C4} & E=\frac 12\mu
\left(\stackrel{.}{A_1}^2+\stackrel{.}{A_2}^2\right)
+V[A_1,A_2]\nonumber\\ & +\frac 12\sum\limits_\alpha \{m_\alpha
\stackrel{.}{\xi }_{\alpha ,1}^2+m_\alpha \stackrel{.}{\xi
}_{\alpha ,2}^2+m_\alpha \omega _\alpha ^2\left( \xi _{\alpha
,1}-\frac{c_\alpha A_1}{m_\alpha \omega _\alpha ^2}\right)
^2\nonumber\\ &+m_\alpha \omega _\alpha ^2\left( \xi _{\alpha
,2}-\frac{c_\alpha A_2}{m_\alpha \omega _\alpha ^2}\right) ^2\}
\end{align}
\noindent In the Euclidean action ($S^{(E)}$) contains
contributions from the system ($S_S^{(E)}$), the reservoir
($S_R^{(E)}$) and the interaction ($S_I^{(E)}$),
\begin{align}
\label{B1115b} S^{(E)}&=S_S^{(E)}+S_R^{(E)}+S_I^{(E)}\nonumber\\
                  &=S_S^{(E)}+\int\limits_0^{\beta \hbar }d\tau (L_R^{(E)}+L_I^{(E)})
\end{align}
\noindent with
\begin{align}
\label{B55a}S_S^{(E)}=& S_{opt}=\int\limits_0^{\beta \hbar
}d\tau\{\frac 12\mu\left( \frac{dA_1(\tau)}{d\tau}\right) ^2-\frac
12\mu\Omega ^2A_1^2(\tau)\nonumber\\&+\frac 12\mu\left(
\frac{dA_2(\tau)}{d\tau}\right) ^2 +\frac 12\mu\omega _2^2(
\overline{A}_2)(A_2(\tau)-\overline{A}_2)^2\nonumber\\
                        &+L_2(\overline{A}_2)\}.
\end{align}
In this optimized action two modes are decoupled. Also
\begin{equation}
\label{B116} L_R^{(E)}=\sum\limits_{\alpha =1}^N \frac 12 m_\alpha
(\overset{.}\xi_{\alpha,1} ^2+\omega _\alpha ^2{{\xi }_{\alpha
,1}}^2+\overset{.}\xi_{\alpha,2} ^2+\omega _\alpha
^2{{\xi}_{\alpha ,2}}^2)
\end{equation}
\begin{equation}
\label{B116a} L_I^{(E)}=\sum\limits_{\alpha =1}^N(-c_\alpha
\xi_{\alpha,1} A_1+\frac 12\frac{c_\alpha ^2A_1^2}{m_\alpha
\omega _\alpha ^2}-c_\alpha \xi_{\alpha,2} A_2+\frac
12\frac{c_\alpha ^2A_2^2}{m_\alpha \omega _\alpha ^2})
\end{equation}
\noindent Now we write Euclidean action for the two modes (with
bath) separately. For the first mode
\begin{align}
\label{B116b}& S_1^{(E)}=\int\limits_0^{\beta \hbar }d\tau\{\frac
12\mu\left( \frac{dA_1(\tau)}{d\tau}\right) ^2-\frac 12\mu\Omega
^2A_1^2(\tau)\nonumber\\& +\sum\limits_{\alpha =1}^N \{\frac 12
m_\alpha (\overset{.}\xi_{\alpha,1} ^2+\omega _\alpha ^2{{\xi
}_{\alpha ,1}}^2)-c_\alpha \xi_{\alpha,1} A_1+\frac
12\frac{c_\alpha ^2A_1^2}{m_\alpha \omega _\alpha ^2}\}\}
\end{align}
and for the second mode
\begin{align}
\label{B116c}& S_2^{(E)}= \int\limits_0^{\beta \hbar }d\tau\{\frac
12\mu\left( \frac{dA_2(\tau)}{d\tau}\right) ^2 \nonumber\\& +\frac
12\mu\omega _2^2(\overline{A}_2)(A_2(\tau)-\overline{A}_2)^2
+L_2(\overline{A}_2)\nonumber\\&+\sum\limits_{\alpha =1}^N \{\frac
12 m_\alpha (\overset{.}\xi_{\alpha,2} ^2+\omega _\alpha ^2{{\xi
}_{\alpha ,2}}^2)-c_\alpha \xi_{\alpha,2} A_2+\frac
12\frac{c_\alpha ^2A_2^2}{m_\alpha \omega _\alpha ^2}\}\}
\end{align}
The expression for $S_1^{(E)}$ is equivalent to the expression for
an inverted parabolic potential coupled to $N$ harmonic
oscillatore. So we  have a system of ($N+1$) harmonic oscillators
and we can calculate $\Omega^{fric}$ using the method discussed in
the classical calculation \cite{Sayan}. Now we will calculate the partition
function using the `action' defined in Eq. (\ref{B116c}).
Following Weiss \cite{Weiss} we define ``influence action''
$S_{R,I}^{(E)}(A_2,\bm{\xi})$, which captures the influence of the
environment on the equilibrium properties of the open system.
\begin{equation}
\label{B116a} S_2^{(E)}= S_{R,I}^{(E)}[A_2,{\bm
\xi}]=S_{R}^{(E)}[{\bm \xi}]+S_{I}^{(E)}[A_2,{\bm \xi}]
\end{equation}
\noindent The stationary paths of action, which we denote by
$\overline{A}_2$ and by $\overline{\xi}_{\alpha,2}$, obey the
Euclidean classical equations of motion
\begin{align}
\label{B117} \mu\overset{..}{\overline{A_2}}-\frac{\partial
V(\overline{A}_2)}{\partial \overline{A_2}}+\sum\limits_{\alpha
=1}^Nc_\alpha (\overline{\xi}_{\alpha,2} -\frac{c_\alpha
\overline{A}_2}{m_\alpha \omega _\alpha ^2})=0 \nonumber\\
m_\alpha \overset{..}{\overline{\xi}}_{\alpha,2} -m_\alpha
\omega _\alpha ^2\overline{\xi}_{\alpha,2} +c_\alpha
\overline{A}_2=0
\end{align}
\noindent We choose for convenience to periodically continue the
paths $\xi_{\alpha,2}(\tau)$, $A_2(\tau)$ outside the range $0
\leq \tau < \beta \hbar $ by writing them as Fourier series
\begin{align}
\label{B118} \xi_{\alpha,2}(\tau )=\sum\limits_{n=-\infty
}^{n=\infty }\xi_{\alpha,2 ,n}e^{i\nu _n\tau
}\nonumber\\
A_2(\tau )=\sum\limits_{n=-\infty }^{n=\infty }A_{2,n}e^{i\nu
_n\tau }
\end{align}
\noindent where $\xi_{\alpha,2,n}=\xi_{\alpha,2,-n}^{*}$,
$A_{2,n}^{*}=A_{2,-n}^{*}$ and $\nu _n=\frac{2\pi n}{\beta \hbar
}$ is a bosonic Matsubara frequency. Substituting Eq. (\ref{B118})
into Eq. (\ref{B116a}) and Eq. (\ref{B116}), we obtain
\begin{align}
\label{B119} S_{R,I}^{(E)}[A_2,{\bm \xi}]=& \sum\limits_{\alpha
=1}^N {\beta \hbar} \sum\limits_{n=-\infty }^{n=\infty
}\frac{m_\alpha }2\nonumber\\&(\nu _n^2\left| \xi_{\alpha,2
,n}\right| ^2+\omega _\alpha ^2\left| \xi_{\alpha,2
,n}-\frac{c_\alpha }{m_\alpha \omega _\alpha ^2}A_{2,n}\right|
^2).
\end{align}
\noindent Next, we decompose $\xi_{\alpha,2,n}$ into classical
term $\overline{\xi}_{\alpha,2,n}$ and a deviation
$y_{\alpha,2,n}$ describing quantum fluctuations,
\begin{align}
\label{B120}
\xi_{\alpha,2,n}&=\overline{\xi}_{\alpha,2,n}+y_{\alpha,2,n}\nonumber\\
            &= \frac{c_\alpha }{m_\alpha (\nu _n ^2+\omega _\alpha
            ^2)}A_{2,n}+y_{\alpha,2,n}.
\end{align}

\begin{figure}
\centering \epsfig{file=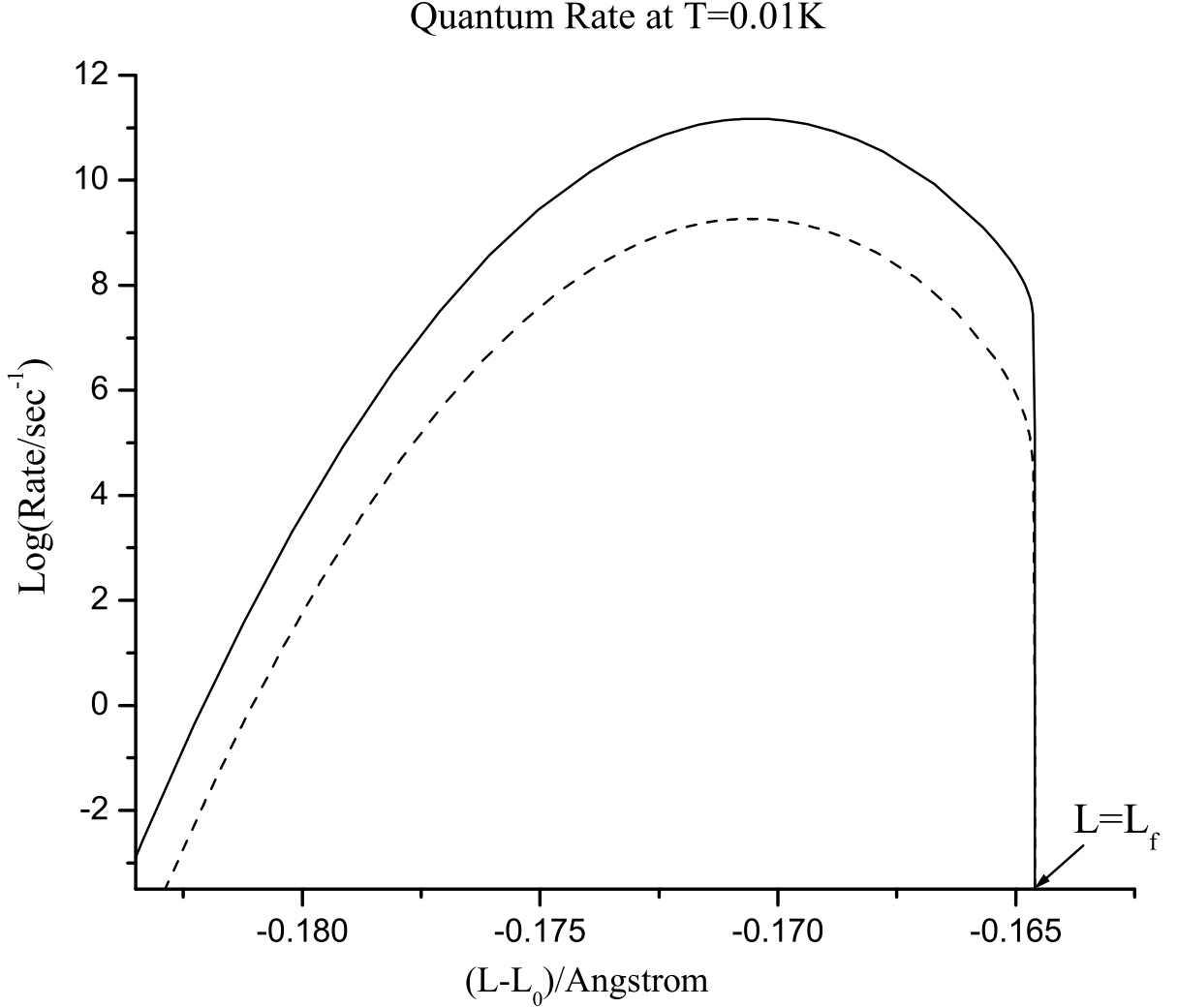,width=1\linewidth}
\caption{Plot of logarithm of rate of crossing from one buckled
state to the other using quantum transition state theory without
friction (solid line) and with friction (dashed line), for a Si
rod of dimensions $L_{0}=500 \:\AA$,$w=20 \:\AA$, $d=10 \:\AA$,
$\gamma=1\times10^{11}sec^{-1}$ and $T=0.01\:K$. For this rod the first three instabilities occur at
$L_{f}-L_{0}=-0.1646\:\AA$, $L_{s}-L_{0}=-0.6597\:\AA$ and $\gamma=1\times10^{11}sec^{-1}$.}
\label{figQfric50T001}
\end{figure}

\noindent In the second form, we have used the solution of the
oscillator mode $\xi_{\alpha,2,n}$ following from Eq.
(\ref{B117}). Since $\xi_{\alpha,2}(\tau)$ is a stationary point
of action, the term linear in the deviation is eliminated and we
find an expression in which the quadratic forms of $\bm{y}$ and
$A_2$ are decoupled,
\begin{equation}
\label{B121} S_{R,I}^{(E)}[A_2,{\bm{\overline{\xi}}+{\bm
y}}]=S_{R,I}^{(E)}[{\bm y}]+S_{\inf l}^{(E)}[A_2],
\end{equation}
\noindent
\begin{align} \label{B122} S_{R,I}^{(E)}[{\bm y}]&=
\sum\limits_{\alpha =1}^N {{\beta \hbar }}\sum\limits_{n=-\infty
}^{n=\infty }\frac{m_\alpha }2(\nu _n^2+\omega _\alpha ^2)\left|
y_{\alpha,2
,n}\right| ^2 \nonumber\\
                             &= \sum\limits_{\alpha =1}^N\int\limits_0^{\beta \hbar }d\tau
\frac{m_\alpha }2 (\overset{.}{y}_{\alpha,2} ^2+\omega _\alpha
^2y_{\alpha,2}^2),
\end{align}
\noindent
\begin{equation}
\label{B123} S_{infl}^{(E)}[A_2]=\sum\limits_{\alpha =1}^N {{\beta
\hbar }}\frac{c_\alpha ^2}{2m_\alpha }\sum\limits_{n=-\infty
}^{n=\infty }\left( \frac{\left| A_{2,n}\right| ^2}{\omega _\alpha
^2}-\frac{\left| A_{2,n}\right| ^2}{(\nu_n ^2+\omega _\alpha
^2)}\right).
\end{equation}
\noindent In the Fourier series representation, the influence
action (Eq. (\ref{B123})) takes the complete form
\begin{equation}
\label{B124} S_{infl}^{(E)}[A_2]={\mu}{\beta \hbar
}\sum\limits_{n=-\infty }^{n=\infty }(\zeta _n\left|
A_{2,n}\right|^2),
\end{equation}
\noindent where
\begin{align}
\label{B125} \zeta_n &= \frac 1{\mu}\sum\limits_{\alpha
=1}^N\frac{c_\alpha ^2}{2m_\alpha }\frac{\nu _n^2}{(\nu
_n ^2+\omega _\alpha ^2)}\nonumber\\
                     & =\frac 2{\mu\pi }
                     \frac 1{\beta \hbar }\int\limits_0^\infty d\omega \frac{J(\omega )}\omega \frac{\nu _n^2}{(\nu _n^2+\omega
                     ^2)}.
\end{align}
\noindent Assuming Ohmic friction \cite{Weiss}
\begin{equation}
\label{B126} \zeta_n=\nu _n\gamma.
\end{equation}
\noindent We want to calculate the ratio of partition functions
for the second mode at the saddle and at the reactant geometry. As
all other modes are harmonic, there contributions to the rate is
calculated using the same procedure as discussed in the section 3.1. In our calculation we use the `action' for the second
mode at the saddle point, given in the Eq. (\ref{B116c}).
Expanding the kinetic term into its Fourier components, we get
\begin{align}
\label{B127}& S_2^{(E)}= \mu\beta \hbar \left[
\sum\limits_{n=1}^\infty (\nu _n^2+\omega _2^2(\overline{A}_2)+\nu
_n\gamma )\left| A_{2,n}\right| ^2+L_2(\overline{A}_2)\right]
\end{align}
\noindent So the partition function  for the second mode at the
saddle point is given by
\begin{align}
\label{B129}Q_2^{\ddagger }=&\int\limits_{-\infty }^\infty
\frac{d\overline{A}_2e^{-\frac{L_2(\overline{A}_2)}{kT}}}{\sqrt{2\pi
\beta \hbar ^2/\mu}}\Gamma [\frac{\beta \hbar }{4\pi }(\frac{4\pi
}{\beta \hbar }+\gamma -\sqrt{\gamma ^2-4\omega
_2^2(\overline{A}_2)})] \nonumber\\& \times \Gamma [\frac{\beta
\hbar }{4\pi }(\frac{4\pi }{\beta \hbar }+\gamma +\sqrt{\gamma
^2-4\omega _2^2(\overline{A}_2)})]\times Q_{bath}.
\end{align}
\noindent We have already determined $\Omega ^2$, $\omega _2^2(\overline{A}_2)$ and $L_2(%
\overline{A}_2)$ variationally in the absence of reservoir. For
the second mode at the reactant geometry, the expression for the
partition function is given below
\begin{align}
\label{B130}Q_2=&\int\limits_{-\infty }^\infty
\frac{d\overline{A}_2e^{-\frac{\mu \omega_2^2
\overline{A}_2^2}{2kT}}}{\sqrt{2\pi \beta \hbar ^2/\mu}}\Gamma
[\frac{\beta \hbar }{4\pi }(\frac{4\pi }{\beta \hbar }+\gamma
-\sqrt{\gamma ^2-4\omega _2^2})] \nonumber\\& \times \Gamma
[\frac{\beta \hbar }{4\pi }(\frac{4\pi }{\beta \hbar }+\gamma
+\sqrt{\gamma ^2-4\omega _2^2})]\times Q_{bath}.
\end{align}
\noindent In the regime where $L > L_s$, the rate may be calculated using (transition state is assumed to be straight rod)
\begin{equation}
\label{B111a} R_{quantum,i}^{fric}=\rho_f (\frac{\omega _1}{2\pi
})(\frac{\Omega^{fric}_{Linear}}{\Omega_{Linear} }){\frac
{Q_2^{\ddagger}}{Q_2}}\prod\limits_{n=3}^\infty \frac{\omega
_n}{\omega_{Linear,n}^{\ddagger }}e^{-\beta \Delta E_{Barrier}^{Linear}}.
\end{equation}
\noindent In the regime where $L < L_s$, the rate may be calculated using (transition state is assumed to be bent rod)
\begin{equation}
\label{B111b} R_{quantum,i}^{fric}=\rho_s (\frac{\omega _1}{2\pi
})(\frac{\Omega^{fric}_{Bent}}{\Omega_{Bent} }){\frac
{Q_2^{\ddagger}}{Q_2}}\prod\limits_{n=3}^\infty \frac{\omega
_n}{\omega_{Bent,n}^{\ddagger }}e^{-\beta \Delta E_{Barrier}^{Bent}}.
\end{equation}
\begin{figure}
\centering \epsfig{file=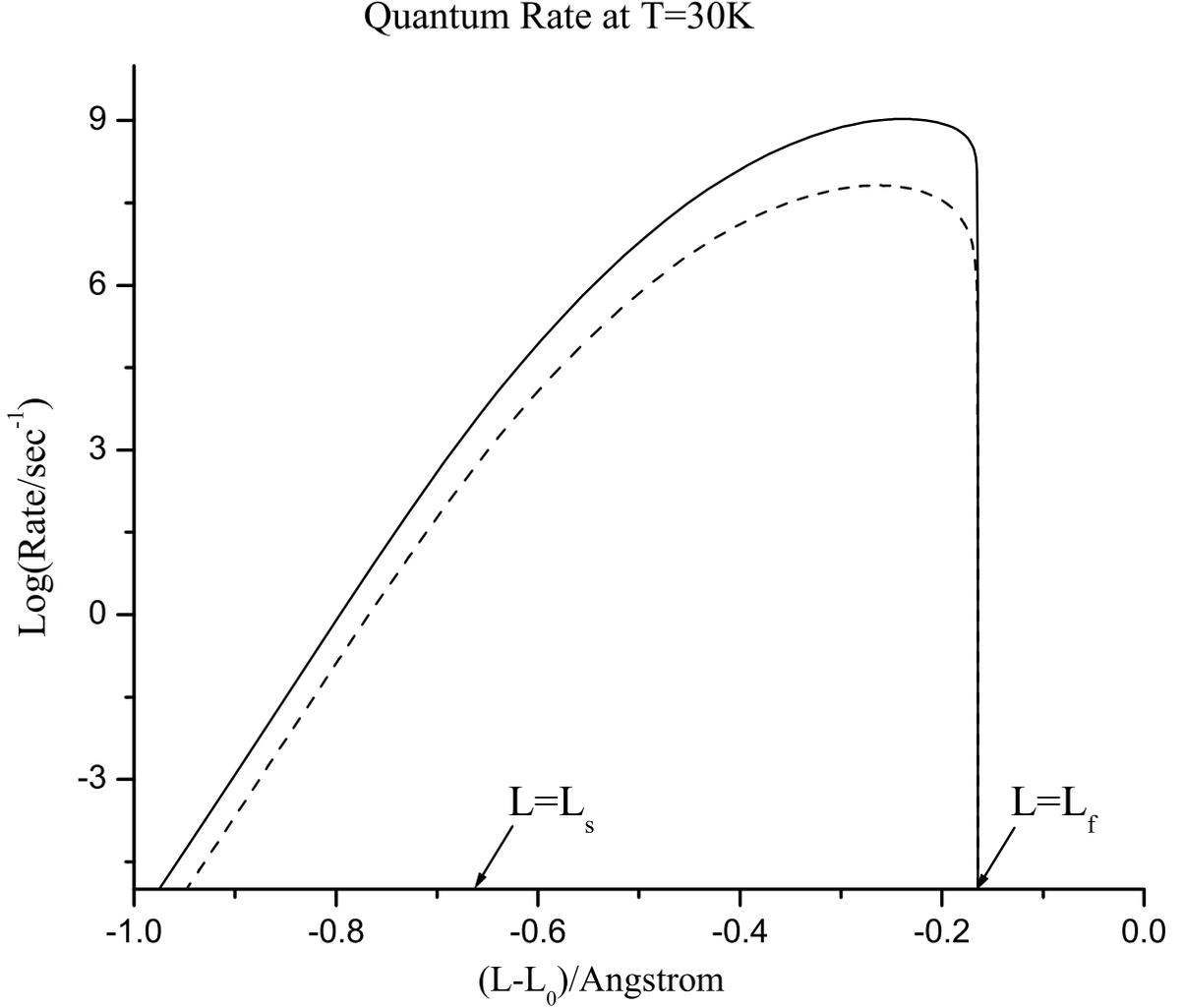,width=1\linewidth}
\caption{Plot of logarithm of rate of crossing from one buckled
state to the other using quantum transition state theory without
friction (solid line) and with friction (dashed
line), for a Si rod of dimensions $L_{0}=500 \:\AA$,$w=20 \:\AA$,
$d=10 \:\AA$ and $T=30\:K$. For this rod the first three instabilities occur at
$L_{f}-L_{0}=-0.1646\:\AA$, $L_{s}-L_{0}=-0.6597\:\AA$ and
 $\gamma=1\times10^{11}sec^{-1}$.} \label{figQfric50T30}
\end{figure}

\section{Results}
\noindent We now use the above formulae to calculate the rate of
passage from one buckled state to the other, for $Si$ rod of dimensions
$L_{0}=500 \:\AA$,$w=20 \:\AA$, $d=10 \:\AA$,  the
case considered by Carr {\it et al.} \cite{Carr}. The Young's modulus and density of $Si$ are
$Q=130\:GPa$ and $\rho =5000\:kg.m^{-3}$.  The first and second buckling
instabilities occur at lengths which we denote as $L_f$ and  $L_s$ 
respectively and their values are given by
$L_{f}-L_{0}=-0.1646\:\AA$, $L_{s}-L_{0}=-0.6597\:\AA$. In
the regime $L \geq L_s$, the saddle point has straight rod
configuration and in the regime $L < L_s$, the saddle point has
bent configuration. At $T=0.01\:K$ we do not show the rate calculated near the
second buckling instability - as it is extremely low (Fig. \ref{figQfric50T001}).  The equations for rates have product over all
the infinite modes, as would be normal for a continuum model. This
however is not realistic, and this can be easily modified to
account for discreteness of these rod. We do this, by restricting
the product to contain just the same number of normal modes as the
perpendicular degrees of freedom for the discrete model. We have
thus taken contributions from first $2128$ normal modes of the
rod. Fig. \ref{figQfric50T30} shows the
quantum rate against compression, made at a temperature of
$30\:K$.  Our calculation including friction shows that friction lowers rate of
conversion from one buckled state to the other.
\noindent

\section{Summary and Conclusions}

\noindent In this paper we have discussed the buckling of a nano
rod under compression and rate of it's conversion from one buckled
state to the other using quantum version of
multidimensional transition state theory. As compression
increases, the rod buckles (first buckling instability), and has
now two stable states. From one stable state it can go over to the
other by thermal fluctuations or quantum tunneling. Using a
continuum approach, we have calculated the rate of conversion from
one state to the other using system plus reservoir model. The saddle
point for the change from one state to the other is the straight
rod configuration.  The rate expression, however, diverges at
the second buckling instability. At this point, the straight rod
configuration, which was a saddle till then, becomes hill top and
two new saddles are generated. The new saddles have bent
configurations and as the rod goes through further instabilities,
they remain stable and the rate calculated according to the
harmonic approximation around the saddle point remains finite.
However, this rate too, diverges near the second buckling
instability. In the quantum transition state theory calculation we have
 calculated centroid partition function for the second mode, and derived expressions that are well
behaved through the second buckling instability. Using these expressions, we have calculated the rates for
nano-rods of dimensions $L_{0}=500 \:\AA$,$w=20 \:\AA$, $d=10 \:\AA$. 

\section*{Acknowledgments}
The author is greatful to Prof. K. L. Sebastian for his continuous guidance and for suggesting this very interesting problem. The author also thanks Prof. Eli Pollak for his comments on the manuscript.

\end{document}